\documentclass[twocolumn,showpacs,prl]{revtex4}

\newcommand{\BE}{\begin{equation}}
\newcommand{\EE}{\end{equation}}
\newcommand{\BA}{\begin{eqnarray}}
\newcommand{\EA}{\end{eqnarray}}
\usepackage{graphicx}% Include figure files
\usepackage{dcolumn}% Align table columns on decimal point
\usepackage{bm}% bold math

\begin{document}

\title{Are waves all localized in two dimensional random media?}

\author{Zhen Ye}\affiliation{Department of Physics, National Central University,
Chungli, Taiwan 32054}

%\date{February 2, 2002}
\date{October 16, 2002}

\begin{abstract}

It has been the dominant view for over two decades that all waves
are localized in two dimensions for any given amount of disorder.
Here, questions are raised about this assertion. It is shown that
there is a lack of the convincing support of the claim. Rather,
the recent evidence tends to indicate that waves are not
necessarily always localized in two dimensional random systems.

\end{abstract}

 \maketitle

The concept of localization was originally introduced by
Anderson\cite{Anderson58} for electrons in a crystal. In the case
of a perfectly periodic lattice, except in the gaps all the
electronic states are extended and are represented by Bloch
states. When a sufficient amount of disorders is added to the
lattice, for example in the form of random potentials, the
electrons may become spatially localized due to the multiple
scattering by the disorders. In such a case, the eigenstates are
exponentially confined in the space. By a scaling
analysis\cite{gang4}, Abraham {\it et al}. suggested that there
can be no metallic state or metal-insulator transition in two
dimensions in zero magnetic field. In other words, all electrons
are always localized in two dimensions (2D), as reviewed in
\cite{Lee,EA}.

The fact that the electronic localization is due to the wave
nature of electrons has led to the conjecture that the
localization phenomenon also exists for classical waves in random
media. And all predictions for the electronic localization are
believed to hold for classical waves.
%Now the general perception
%of localization is: It refers to situations in which due to
%multiple scattering by disorders waves are confined in space, and
%remain trapped until dissipated; meanwhile the energy density
%decays exponentially from the transmitting point.
Following the scaling analysis of the electronic localization, it
was widely accepted that all waves are localized in 2D random
media. This has been the prevailing view for the past twenty years
(e.~g. Ref.~\cite{Condat,McCall,Sigalas}). Hereafter I will refer
to this view as `2D Conjecture'

In this Letter, I propose that the popular view on 2D localization
may not be valid. While the `2D Conjecture' has been challenged
and shown to be likely incorrect for electronic systems (Reviewed
in \cite{EA}), here I will focus on classical waves. For the
purpose, I will first review the origin of the `2D Conjecture' and
the current theory on 2D wave localization. Then point out the
ambiguities in the theory and discuss the evidence that is in
conflict with the `2D Conjecture'. We take the following steps.
(1) Re-inspect the scaling analysis, checking for ambiguities. (2)
Examine the predictions from the current theory, checking for its
validity. (3) Examine the previous experimental and numerical
results that claim to support the `2D Conjecture', checking for
their appropriateness. (4) Find the self-conflicting points in the
current theory, and discuss an apparent mechanism which is in
conflict with the theory. The idea is that if there are conflicts,
the 2D conjecture should be at least skeptical. The confidence is:
Wave scattering in many 2D systems is exactly calculable following
Twersky\cite{Victor}; thus previous predictions can be put under a
close scrutiny. We note that the electronic system is more
complicated because effects such as the Coulomb interaction makes
data interpretation difficult. In this sense, the classical wave
systems are advantageous in studying localization effects.

\input{epsf}
\begin{figure}[hbt]
\begin{center}
\epsfxsize=2.25in \epsffile{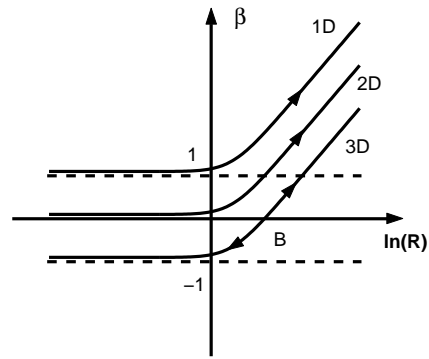} \caption{ \label{fig1}\small
The scaling function $\beta$ versus  $\ln R$ from
Eq.~(\ref{eq:4})}
\end{center}
\end{figure}

{\bf The scaling analysis} According to \cite{gang4}, an
hypercubic geometry is used for the scaling analysis. In the
metallic state, the resistance follows the Ohmic behavior \BE R
\sim L^{2-d},\label{eq:1}\EE where $d$ is the dimension. For a
localized state, i.~e. large $R$, the resistance grows
exponentially \BE R\sim e^{L/L_1},\label{eq:2}\EE where $L_1$ is
the localization length which may differ for different dimensions.
A scaling function is defined as $\beta = \frac{\partial \ln
R}{\partial \ln L}.$ Taking Eqs.~(\ref{eq:1}) and (\ref{eq:2})
into this, we obtain the asymptotic behavior \BE \beta \sim
\left\{
\begin{array}{ll} \ln R, & \mbox{as} \ R\rightarrow \infty \
(\mbox{Localized})
\\ 2-d, & \mbox{as} \ R \rightarrow 0 \ (\mbox{Ohmic})
\end{array}\right. \label{eq:4}
\EE From the asymptotic behavior in Eq.~(\ref{eq:4}), one can
sketch the universal curves in $d=1,2,3$ dimensions. The central
assumptions in \cite{gang4} are (1) $\beta$ is continuous; (2)
$\beta$ is a function of $R$ and depends on other parameters such
as disorders and length scale only through $R$; and (3) once wave
is localized, the increasing sample size would always mean more
localization.

The behavior of $\beta$ is plotted in Fig.~\ref{fig1}. It is clear
that in the 3D case, the curve crosses the horizontal axis,
yielding an unstable fixed point ($B$). Above this point, the
waves become more and more localized as the sample size increases.
Below the critical point, the system tends to follow the Ohmic
behavior as the sample size is enlarged. This fixed point
separates the localized and non-localized states. For the two
dimensional case, in the Ohmic regime $\beta$ approaches zero as
$\ln(R) \rightarrow 0$. But the perturbation calculation including
the wave interference effect shows that $\beta$ is always greater
than zero. Therefore for both one and two dimensions, the curves
do not cross the horizontal axis, and there is thus no fixed
point. As the sample size increases, all states move towards the
localization regime. {\it This has been the main reason that led
previously to the conclusion that all waves are localized in one
and two dimensions.}

{\bf The current localization theory} Now I briefly review the
existing theory for localization. As wave propagates in random
media, it experiences multiple scattering, and as a result, the
wave loses its phase, leading to the gradual decreases of the
coherence of the wave in the absence of absorption. Meanwhile,
diffusive wave is built up as more and more scattering takes
place. The procedure to obtain the localization state can be
briefly summarized as follows.

The quantity, $D^{(B)}$ which is a measure of diffusion of
classical waves is called the classical Boltzman diffusion
constant and it may be derived under the coherent potential
approximation \cite{lax}, and is given as
\begin{equation}
D^{(B)} \sim \frac{v_t l}{d}
\end{equation}
where $v_t$ is the transport velocity, $l$ is the mean free path
and $d$ is the dimensionality.

As waves scattered along any two reversed paths in the backward
direction interfere constructively, leading to the enhanced
backscattering effect, which will add corrections to the diffusion
coefficient. In the field theory approach, such an enhanced
backscattering effect is represented by a set of maximally crossed
ladder diagrams\cite{DV}. In the two dimension case, the
evaluation of these diagrams leads to an integration for which two
cut-off limits have to be introduced to avoid the divergence. The
correction to the diffusion constant for two dimensional systems
is thus found as
\begin{equation}
\delta D \sim  - {\rm ln}(L_M/l_m) \label{eq:correction}
\end{equation}
where $L_M$ and $l_m$ are the two cut-off limits. It is then
interpreted in the previous theory that the cut-off limit $l_m$ is
a measure of the minimum scaling for the waves and is thought to
be related to (for example) the mean free path, whereas $L_M$ is a
measure of the effective size of the sample. It is rather
important to note that the correction in Eq.~(\ref{eq:correction})
is not only negative but diverges as $L_M$ approach infinity. This
is obviously unphysical, since the corrected diffusion constant
cannot be negative. To avoid the problem, it was suggested that
$L_M$ is related to the localization range, or simply the
localization length, in such way that when $L_M$ is equal to the
localization length denoted by $\xi$ say, the corrected diffusion
coefficient becomes zero: \BE D_R(\xi) = D^{(B)} + \delta D(\xi) =
0. \label{eq:d} \EE The localization length $\xi$ is subsequently
solved for from this equation. It is obvious that this equation
{\it always} allows a solution. {\it Therefore a localization
length can always be found in two dimensions.} {\it Such a
backscattering induced absence-of-diffusion mechanism is the core
of the current theory of localization in two dimensions, and is
considered a strong support of the `2D Conjecture' from the
scaling analysis}\cite{gang4}.

{\bf On the scaling analysis} While it is simple and
straightforward, the above scaling analysis is not without
ambiguities in investigating the localization effect. The reasons
follow. Whether a system has non-localized or only localized
states is an intrinsic property of the system, and should not rely
on neither the boundary nor the source. As long as the analysis
cannot exclude the possibility that the boundary or the source is
playing a role, the consequence from the analysis is deemed to be
questionable. In order to isolate the localization or
non-localization effect, therefore, a genuine analysis should not
be plagued by boundary effects not only in the localization region
but also in the non-localization region. Of course, if the system
has indeed only localized states, the boundary is not an issue, as
the dependence on the boundary is exponentially vanishing.
However, the care must be taken for the non-localized regime. It
is not difficult to see that the above scaling theory may work for
situations when both probing contacts, used to measure the
resistance or conductance from which the localization is inferred,
are located outside the system. In this case, the Ohmic behavior
given by Eq.~(\ref{eq:1}) is valid under the condition that the
current flows uniformly in one direction. This is possible only
with properly scaled sources and with the presence of confining
boundaries, obviously in conflict with the proclamation that
whether it is a localization or non-localized state is the
intrinsic property of the system and should not rely on a boundary
nor a source. Thus the above analysis is more appropriate for
studying transport phenomena. It is our opinion that the reduction
in the conductance does not necessarily mean that all waves are
actually localized. In other words, it is necessary to
differentiate the situation that the electrons are prohibited from
transmission through a random medium from the situation that the
system has only localized states.

Not to be neglected, the recent numerical results on acoustic
waves\cite{PRE2002} do support the above comments on the scaling
theory. They point out that waves can be blocked from propagation
{\it through} a medium by disorders, but such an inhibition is not
necessarily an indication that waves can be localized in the
medium when the transmitting source is put {\it inside} the
medium. The same conclusion is also obtained from the simulations
on propagation of electromagnetic waves (EM) in 2D random
dielectric media\cite{Bikash2}. More explicitly, it is shown that
waves certainly cannot propagate through a random medium when it
has only localized states. But the observation that waves cannot
propagate through does not necessarily imply that the system has
only localized states.

To this end, it is proper to mention that a recent scaling
analysis shows that the difference between 2D and 3D disordered
systems is insignificant\cite{scaling}, i.~e. like 3D the
localization-delocalization transition is also possible in 2D,
which could explain the recently observed `unusual' metallic
behavior in 2D electronic systems (See references cited in
\cite{EA}). This scaling analysis has been pointed out to be
absolutely in line with recent findings obtained through direct
calculation of the conductance with the use of the Kubo formula
(Y. Tarasov, {\it private communication}).

{\bf On the validity of the current theory} Among many, there are
two basic ways to check the validity of the aforementioned theory.

\begin{figure}[hbt]
\begin{center}
\epsfxsize=2.25in \epsffile{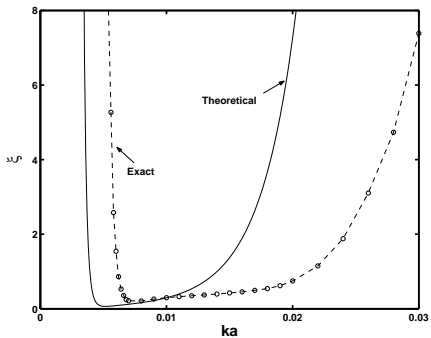} \caption{Localization length
($\xi$) is shown as a function of frequency ($ka$) for
$\beta$=0.001. The dashed curve with circles represent the exact
values obtained numerically while the solid curve is obtained from
theory. Here $k$ is the wavenumber, $a$ is the radius of the
cylinder, and $\beta$ is the fraction of area occupied by the
cylinders per unit area.} \label{fig2}
\end{center}
\end{figure}

First, the key parameter obtained from the current theory is the
localization length. This quantity can also be obtained exactly by
numerical computation using the scheme detailed in\cite{Victor}.
Then we can make a comparison of the two results. We consider the
model of acoustic scattering in water with air-cylinders detailed
in \cite{Emile}. The comparison is shown in Fig.~\ref{fig2}. The
comparison clearly shows that there is a significant difference
between the numerically exact results and the results obtained
from the theory. Further comparisons indicate that the difference
between the two results is significant not only quantitatively but
qualitatively\cite{Bikash3}.

\input{epsf}

Second, one can evaluate the localization length from the theory.
Using an exactly solvable model, then inspect the spatial
distribution of the wave energy density and check whether
localization occurs when the system is larger than the obtained
localization length. Again we consider the model of acoustic
localization in air-filled cylinders in water. From
Fig.~\ref{fig2}, the theory predicts that the shortest
localization length is around $ka=0.005$ which is the nature
frequency of an air-cylinder in water\cite{Emile}. The numerical
results show, however, that no localization occurs around this
frequency. Instead, localization appears at somewhat higher
frequency ranges with the same system size, referring to Fig.~3 of
\cite{Emile}.

{\bf On the previous experimental and numerical evidence} It was
widely believed that the theory of 2D localization has been tested
experimentally to be successful. We find that the claimed success
is mainly based on two types of experiments. One is the indirect
method which measures the effects of the enhanced backscattering
(e.~g. Ref.~\cite{2Dexp}). In a rigorous simulation, it has been
shown that the enhanced backscattering is not related to the
localization\cite{aad}. Waves are not necessarily localized when a
strong enhanced backscattering exits, and sometimes waves can be
localized even when the enhanced backscattering is weak. Asides
from few exceptions\cite{McCall}, the other type of experiments is
based on observations of the exponential decay of waves as they
propagate {\it through} disordered media, as stated in
Ref.~\cite{Nat}. According to the above and following discussions,
this type of experiments is {\it not} sufficient to discern
whether the medium really only has localized states. Unwanted
effects of non-localization origin can also contribute to the
wanted exponential decay, making data interpretation ambiguous. In
fact, the authors in Ref.~\cite{Sigalas} pointed out that there is
no conclusive experimental evidence for localization of EM waves
in 2D. We mention that there was a report of the observation of
microwave localization in two dimensions when a transmitting
source is inside disordered media\cite{McCall}. However, the
diffusion based theory has not been verified against this
experimental result.

\begin{center}
\begin{figure}[hbt]
\epsfxsize=2.5in\epsffile{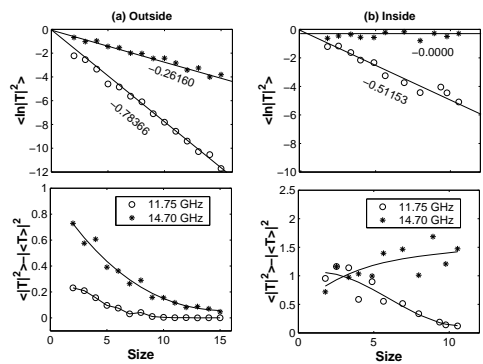} \caption{The logarithmic
average transmission T and its fluctuation versus the sample size
for two frequencies. The estimated slops for the transmission are
indicated in the figure. The `Outside' and `Inside' cases refer
respectively to when the transmitting source is located outside
and inside the medium. More details on the notations and the model
are in \cite{Bikash2}} \label{fig3}
\end{figure}
\end{center}

The same situation can be said about the numerical simulations.
Take Ref.~\cite{Sigalas} as the example. The authors considered
the EM propagation in a random array of dielectric cylinders. The
localization length is computed from the reduction in the
transmission {\it across} the random sample. Following their
method, we first compute the transmission versus the sample size
at various frequencies with the source and the receiver located at
the opposite sides of the sample. We do observe exponential
decays, and the decay rates depend on the frequency. According to
\cite{Sigalas}, the localization lengths can be estimated from
this decay rates. Then as long as the sample size is bigger than
this length, we would also expect to observe an exponential decay
in the transmission when source is put inside the medium. But we
found that the exponential decay disappears for some frequencies.
As an example, the results for two frequencies are shown in
Fig.~\ref{fig3}. In line with \cite{Sigalas}, the following
parameters are used in the computation. The dielectric constants
of the cylinders and the medium are 10 and 1 respectively. The
fraction of area occupied by the cylinders per unit area, is 0.28.
The radius $a$ of the cylinders is 0.38 cm. The lattice constant
$d$ of the corresponding square lattice array is calculated as
1.28 cm. All lengths are scaled by the lattice constant $d$. Here
we see that the exponential decay at 14.70GHz, shown when the
transmission is across the sample, disappears when the source is
moved into the medium. The results suggest that waves are not
localized at this frequency. One may still argue that the
non-localization is due to the fact that the localization length
is long compared to the sample size in the `Inside' case. Even if
this were the case, the exponential decay shown for the `Outside'
case could not be due to the localization effect. Reiterating, it
is not sufficient to extract the localization effect by merely
computing the transmission reduction across the sample. Therefore
the claim about the 2D localization like in \cite{Sigalas} is not
appropriate. At 11.75GHz, we observe that the exponential decay
with nearly the same slop holds for both cases - the slight
difference is due to the finite width in the `Outside'
case\cite{PRE2002}, and reveals the genuine localization behavior.

\input epsf.tex

{\bf The physical picture of localization} It seems that a general
picture of localization may be obtained. For quantum mechanic or
acoustic waves (the same argument also holds for EM
systames\cite{PRER}), the current can be written as $ \vec{J} \sim
\mbox{Re}[\psi(-i)\nabla\psi],$ where $\psi$ stands for the wave
function for quantum mechanical systems and for the pressure in
acoustic systems. Writing the field as $ \psi = |\psi|
e^{i\theta},$ the current becomes $\vec{J} \sim
|\psi|^2\nabla\theta.$ It is clear that when $\theta$ is constant
at least by domains while $|\psi| \neq 0$, the flow stops, i.~e.
$\vec{J}=0$, and the wave or the energy is localized in space,
i.~e. $|\psi|^2 \neq 0$. Obviously the constant phase $\theta$
indicates the appearance of a coherence in the system. This
coherent-phase picture has been demonstrated successfully not only
for two dimensional media\cite{Emile}, but for one and three
dimensions as well\cite{Luan,Hsu}.

The current diffusion-based theory does not support the above
picture. The physical picture of the theory is: Waves will undergo
a diffusion process when the system is smaller than the
localization size. As the system increases, the diffusion
gradually diminishes and finally comes to a complete stop when the
size exceeds the localization length. If this picture were valid,
then one would expect a significant change in the spatial
distribution of the energy density, from that of diffusion to the
exponentially confined envelop. This is not evident in the theory
and is not supported by numerical results. In contrast, the
numerical results in Fig.~\ref{fig3} show that the exponential
decay starts even when the sample size is smaller than the
localization length; the distribution of the characters of a
diffusion process does not appear.

In summary, concerns have been raised about the previous claim
that all waves are localized in 2D. In fact, non-localized states
have already been reported by numerical computation in, for
instance, Ref.~\cite{Emile}. Even if this could be argued to be
due to the finite sample size limited by computing facilities,
there are still many other reasons for being doubtful about the
`2D Conjecture'. Some important reasons are corroborated here.

\end{document}